\input{aipcheck}

\documentclass[
    ,final            
  ]
  {aipproc}

\layoutstyle{8x11single}

\begin{document}

\title{Results from KamLAND-Zen}


\author{K.~Asakura}{
  address={Research Center for Neutrino Science, Tohoku University, Aramaki Aoba, Aoba, Sendai, Miyagi 980-8578, Japan}
}

\author{A.~Gando}{
  address={Research Center for Neutrino Science, Tohoku University, Aramaki Aoba, Aoba, Sendai, Miyagi 980-8578, Japan}
}

\author{Y.~Gando}{
  address={Research Center for Neutrino Science, Tohoku University, Aramaki Aoba, Aoba, Sendai, Miyagi 980-8578, Japan}
}

\author{T.~Hachiya}{
  address={Research Center for Neutrino Science, Tohoku University, Aramaki Aoba, Aoba, Sendai, Miyagi 980-8578, Japan}
}

\author{S.~Hayashida}{
  address={Research Center for Neutrino Science, Tohoku University, Aramaki Aoba, Aoba, Sendai, Miyagi 980-8578, Japan}
}

\author{H.~Ikeda}{
  address={Research Center for Neutrino Science, Tohoku University, Aramaki Aoba, Aoba, Sendai, Miyagi 980-8578, Japan}
}

\author{K.~Inoue}{
  address={Research Center for Neutrino Science, Tohoku University, Aramaki Aoba, Aoba, Sendai, Miyagi 980-8578, Japan}
  ,altaddress={Kavli Institute for the Physics and Mathematics of the Universe (WPI), University of Tokyo, Kashiwa, 277-8583, Japan}
}

\author{K.~Ishidoshiro}{
  address={Research Center for Neutrino Science, Tohoku University, Aramaki Aoba, Aoba, Sendai, Miyagi 980-8578, Japan}
}

\author{T.~Ishikawa}{
  address={Research Center for Neutrino Science, Tohoku University, Aramaki Aoba, Aoba, Sendai, Miyagi 980-8578, Japan}
}

\author{S.~Ishio}{
  address={Research Center for Neutrino Science, Tohoku University, Aramaki Aoba, Aoba, Sendai, Miyagi 980-8578, Japan}
}

\author{M.~Koga}{
  address={Research Center for Neutrino Science, Tohoku University, Aramaki Aoba, Aoba, Sendai, Miyagi 980-8578, Japan}
  ,altaddress={Kavli Institute for the Physics and Mathematics of the Universe (WPI), University of Tokyo, Kashiwa, 277-8583, Japan}
}

\author{R.~Matsuda}{
  address={Research Center for Neutrino Science, Tohoku University, Aramaki Aoba, Aoba, Sendai, Miyagi 980-8578, Japan}
}

\author{S.~Matsuda}{
  address={Research Center for Neutrino Science, Tohoku University, Aramaki Aoba, Aoba, Sendai, Miyagi 980-8578, Japan}
}

\author{T.~Mitsui}{
  address={Research Center for Neutrino Science, Tohoku University, Aramaki Aoba, Aoba, Sendai, Miyagi 980-8578, Japan}
}

\author{D.~Motoki}{
  address={Research Center for Neutrino Science, Tohoku University, Aramaki Aoba, Aoba, Sendai, Miyagi 980-8578, Japan}
}

\author{K.~Nakamura}{
  address={Research Center for Neutrino Science, Tohoku University, Aramaki Aoba, Aoba, Sendai, Miyagi 980-8578, Japan}
  ,altaddress={Kavli Institute for the Physics and Mathematics of the Universe (WPI), University of Tokyo, Kashiwa, 277-8583, Japan}
}

\author{S.~Obara}{
  address={Research Center for Neutrino Science, Tohoku University, Aramaki Aoba, Aoba, Sendai, Miyagi 980-8578, Japan}
}

\author{Y.~Oki}{
  address={Research Center for Neutrino Science, Tohoku University, Aramaki Aoba, Aoba, Sendai, Miyagi 980-8578, Japan}
}

\author{M.~Otani}{
  address={Research Center for Neutrino Science, Tohoku University, Aramaki Aoba, Aoba, Sendai, Miyagi 980-8578, Japan}
 ,altaddress={Current address: KEK, High Energy Accelerator Research Organization, 1-1, Oho, Tsukuba, Ibaraki, 305-0801, Japan}
}

\author{T.~Oura}{
  address={Research Center for Neutrino Science, Tohoku University, Aramaki Aoba, Aoba, Sendai, Miyagi 980-8578, Japan}
}

\author{I.~Shimizu}{
  address={Research Center for Neutrino Science, Tohoku University, Aramaki Aoba, Aoba, Sendai, Miyagi 980-8578, Japan}
}

\author{Y.~Shirahata}{
  address={Research Center for Neutrino Science, Tohoku University, Aramaki Aoba, Aoba, Sendai, Miyagi 980-8578, Japan}
}

\author{J.~Shirai}{
  address={Research Center for Neutrino Science, Tohoku University, Aramaki Aoba, Aoba, Sendai, Miyagi 980-8578, Japan}
}

\author{A.~Suzuki}{
  address={Research Center for Neutrino Science, Tohoku University, Aramaki Aoba, Aoba, Sendai, Miyagi 980-8578, Japan}
}

\author{H.~Tachibana}{
  address={Research Center for Neutrino Science, Tohoku University, Aramaki Aoba, Aoba, Sendai, Miyagi 980-8578, Japan}
}

\author{K.~Tamae}{
  address={Research Center for Neutrino Science, Tohoku University, Aramaki Aoba, Aoba, Sendai, Miyagi 980-8578, Japan}
}

\author{K.~Ueshima}{
  address={Research Center for Neutrino Science, Tohoku University, Aramaki Aoba, Aoba, Sendai, Miyagi 980-8578, Japan}
}

\author{H.~Watanabe}{
  address={Research Center for Neutrino Science, Tohoku University, Aramaki Aoba, Aoba, Sendai, Miyagi 980-8578, Japan}
}

\author{B.D.~Xu}{
  address={Research Center for Neutrino Science, Tohoku University, Aramaki Aoba, Aoba, Sendai, Miyagi 980-8578, Japan}
}

\author{Y.~Yamauchi}{
  address={Research Center for Neutrino Science, Tohoku University, Aramaki Aoba, Aoba, Sendai, Miyagi 980-8578, Japan}
}

\author{H.~Yoshida}{
  address={Research Center for Neutrino Science, Tohoku University, Aramaki Aoba, Aoba, Sendai, Miyagi 980-8578, Japan}
 ,altaddress={Current address: Graduate School of Science, Osaka University, Toyonaka, Osaka 560-0043, Japan}
}

\author{A.~Kozlov}{
  address={Kavli Institute for the Physics and Mathematics of the Universe (WPI), University of Tokyo, Kashiwa, 277-8583, Japan}
}

\author{Y.~Takemoto}{
  address={Kavli Institute for the Physics and Mathematics of the Universe (WPI), University of Tokyo, Kashiwa, 277-8583, Japan}
}

\author{S.~Yoshida}{
  address={Graduate School of Science, Osaka University, Toyonaka, Osaka 560-0043, Japan}
}

\author{K.~Fushimi}{
  address={Faculty of Integrated Arts and Science, University of Tokushima, Tokushima, 770-8502, Japan}
}

\author{T.I.~Banks}{
  address={Physics Department, University of California, Berkeley, and Lawrence Berkeley National Laboratory, Berkeley, California 94720, USA}
}

\author{S.J.~Freedman}{
  address={Physics Department, University of California, Berkeley, and Lawrence Berkeley National Laboratory, Berkeley, California 94720, USA}
  ,altaddress={Kavli Institute for the Physics and Mathematics of the Universe (WPI), University of Tokyo, Kashiwa, 277-8583, Japan}
  ,altaddress={Deceased}
}

\author{B.K.~Fujikawa}{
  address={Physics Department, University of California, Berkeley, and Lawrence Berkeley National Laboratory, Berkeley, California 94720, USA}
  ,altaddress={Kavli Institute for the Physics and Mathematics of the Universe (WPI), University of Tokyo, Kashiwa, 277-8583, Japan}
}

\author{T.~O'Donnell}{
  address={Physics Department, University of California, Berkeley, and Lawrence Berkeley National Laboratory, Berkeley, California 94720, USA}
}

\author{L.A.~Winslow}{
  address={Department of Physics and Astronomy, University of California, Los Angels, Los Angels, California 90095, USA}
}

\author{B.E.~Berger}{
  address={Department of Physics, Colorado State University, Fort Collins, Colorado 80523, USA}
}

\author{Y.~Efremenko}{
  address={Department of Physics and Astronomy, University of Tennessee, Knoxville, Tennessee 37996, USA}
  ,altaddress={Kavli Institute for the Physics and Mathematics of the Universe (WPI), University of Tokyo, Kashiwa, 277-8583, Japan}
}

\author{H.J.~Karwowski}{
  address={Triangle Universities Nuclear Laboratory, Durham, North Carolina 27708, USA and Physics Departments at Duke University, North Carolina Central University, and the University of North Carolina at Chapel Hill}
}

\author{D.M.~Markoff}{
  address={Triangle Universities Nuclear Laboratory, Durham, North Carolina 27708, USA and Physics Departments at Duke University, North Carolina Central University, and the University of North Carolina at Chapel Hill}
}

\author{W.~Tornow}{
  address={Triangle Universities Nuclear Laboratory, Durham, North Carolina 27708, USA and Physics Departments at Duke University, North Carolina Central University, and the University of North Carolina at Chapel Hill}
  ,altaddress={Kavli Institute for the Physics and Mathematics of the Universe (WPI), University of Tokyo, Kashiwa, 277-8583, Japan}
}

\author{J.A.~Detwiler}{
  address={Center for Experimental Nuclear Physics and Astrophysics, University of Washington, Seattle, Washington 98195, USA}
}

\author{S.~Enomoto}{
  address={Center for Experimental Nuclear Physics and Astrophysics, University of Washington, Seattle, Washington 98195, USA}
  ,altaddress={Kavli Institute for the Physics and Mathematics of the Universe (WPI), University of Tokyo, Kashiwa, 277-8583, Japan}
}

\author{M.P.~Decowski}{
  address={Nikhef and the University of Amsterdam, Science Park, Amsterdam, the Netherlands}
  ,altaddress={Kavli Institute for the Physics and Mathematics of the Universe (WPI), University of Tokyo, Kashiwa, 277-8583, Japan}
}

\begin{abstract}
KamLAND-Zen reports on a preliminary search for neutrinoless double-beta decay with $^{136}$Xe based on 114.8 live-days after the purification of the xenon loaded liquid scintillator. In this data, the problematic $^{110m}$Ag background peak identified in previous searches is reduced by more than a factor of 10. By combining the KamLAND-Zen pre- and post-purification data, we obtain a preliminary lower limit on the $0\nu\beta\beta$ decay half-life of $T_{1/2}^{0\nu} > 2.6 \times 10^{25}$\,yr at 90\% C.L. The search sensitivity will be enhanced with additional low background data after the purification. Prospects for further improvements with future KamLAND-Zen upgrades are also presented.
%
%
\end{abstract}

\maketitle


\begin{center}
\end{center}

\section{Introduction}

Double-beta ($\beta\beta$) decay, the rarest process that has been experimentally verified so far, is subject to study in both nuclear and particle physics. Unlike charged particles, electrically neutral neutrinos can be Majorana particles, particles that are their own anti-particles, and hence can introduce lepton number non-conservation. Observation of $\beta\beta$ decay without the emission of neutrinos ($0\nu\beta\beta$) would experimentally demonstrate that lepton number is not conserved, and reveal the Majorana nature of neutrinos. Moreover, if the decay process is mediated by the exchange of a light Majorana neutrino, its decay rate is proportional to the square of the effective Majorana neutrino mass $\left<m_{\beta\beta}\right> \equiv \left| \Sigma_{i} U_{ei}^{2}m_{\nu_{i}} \right|$, and therefore its determination would provide a measure of the absolute neutrino mass scale.

\begin{figure}[b]
  \includegraphics[width=0.65\columnwidth]{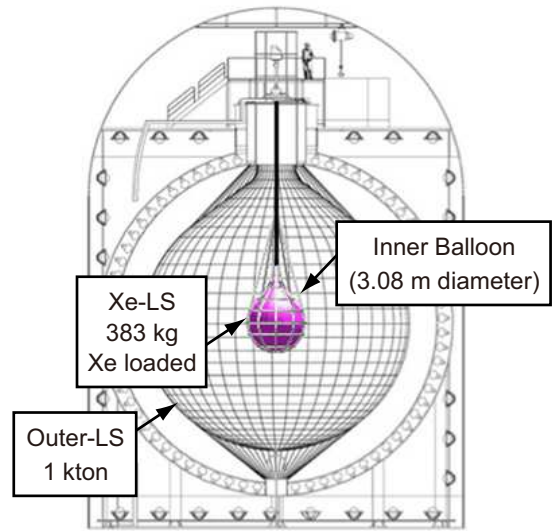}
  \caption{Schematic view of the KamLAND-Zen detector. Phase-1 had 320 kg of $^{136}$Xe enriched Xe dissolved, phase-2 had 383\,kg.}
  \label{figure:detector}
\end{figure}

The KamLAND-Zen $\beta\beta$ decay search experiment started in 2011 and is illustrated schematically in Fig.~\ref{figure:detector}~\cite{Gando2012a}. As the $\beta\beta$ decay source, 320\,kg of $^{136}$Xe enriched xenon gas was dissolved in the liquid scintillator (LS), contained in a transparent nylon balloon (mini-balloon). An initial $0\nu\beta\beta$ decay search with high sensitivity was quickly realized, owing to the extremely low radioacitivity in the already existing KamLAND detector, and the minimization of additional radioacitivities achieved in the manufacturing of the mini-balloon. Based on an initial 213.4 days of measurement (denoted as ``phase-1''), we set a lower limit on the $0\nu\beta\beta$ decay half-life of $T_{1/2}^{0\nu} > 1.9 \times 10^{25}$\,yr at 90\% C.L.~\cite{Gando2013}.

\section{Background}

The $0\nu\beta\beta$ decay search sensitivity in phase-1 was limited by an identified background peak from metastable $^{110m}$Ag. In order to remove this isotope, we embarked on a purification campaign aiming at the reduction of $^{110m}$Ag by a significant factor. In June 2012, we first extracted Xe from the detector, and confirmed that $^{110m}$Ag remained in the Xe-depleted LS. During this process, a diaphragm pump dedicated to the Xe-LS extraction leaked introducing radioactive environmental impurities into the circulating LS. This resulted in an accumulation of radioactive particulate at the bottom part of the mini-balloon. In the meantime, the extracted Xe and additional newly prepared Xe were purified by distillation and adsorption by a getter material. The LS was purified through water extraction and distillation. The purification was expected to be effective, however, we found that $^{110m}$Ag was reduced by only a factor of 3-4, possibly due to $^{110m}$Ag release from the mini-balloon film or partial convection between the original LS and the purified LS in the mini-balloon during filling. We took therefore extra time for the LS purification by three volume exchanges in circulation mode. The processed Xe was dissolved again into the newly purified LS in November 2013. In December 2013, we started the phase-2 data-taking, and found a reduction of $^{110m}$Ag by more than a factor of 10.

After the phase-1 data-taking, we made several efforts for further improvements: (i) the removal of radioactive impurities by Xe-LS purification as mentioned above; (ii) increasing the Xe concentration from $(2.44 \pm 0.01)$~wt\% to $(2.96 \pm 0.01)$~wt\%, indicating the $\beta\beta$ target increase relative to radioactive backgrounds; (iii) developing a spallation background rejection method for $^{10}$C ($\beta^{+}$, \mbox{$\tau = 27.8$~s}, \mbox{$Q = 3.65$~MeV}) from muon-spallation; (iv) optimization of the volume selection to minimize the effect of the mini-balloon backgrounds.

\begin{figure}
  \includegraphics[width=0.7\columnwidth]{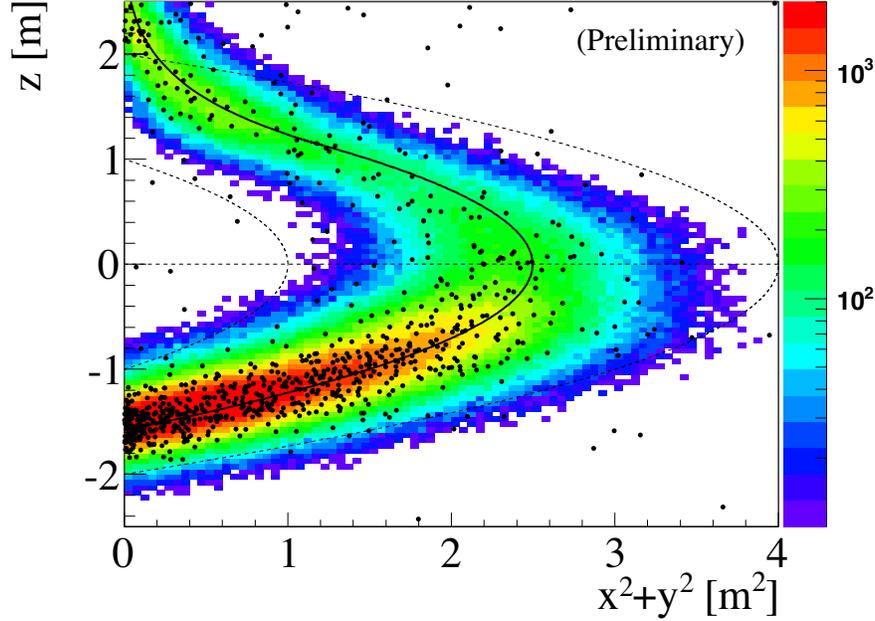}
  \vspace{-2.0cm}
  \caption{Vertex distribution of candidate events (black points) and expected $^{214}$Bi background events from a MC simulation (color histogram) for $2.3 < E < 2.7\,{\rm MeV}$. The normalization of the MC event histogram is arbitrary. The solid line indicates the shape of the balloon film.}
  \label{figure:vertex}
\end{figure}

In the phase-1 data, we observed energy peaks consistent with $^{110m}$Ag background throughout the entire Xe-LS volume and around the mini-balloon, indicating a uniform distribution of $^{110m}$Ag in the Xe-LS, and also on the mini-balloon. The contributions from the Xe-LS and the mini-balloon were almost the same. By contrast, in the phase-2 data, those peaks disappeared, and at present, the primary backgrounds for the $0\nu\beta\beta$ decay search are $^{214}$Bi (daughter of $^{238}$U) on the mini-balloon, the $^{10}$C muon spallation product, and a small contribution from remaining $^{110m}$Ag.

Fig.~\ref{figure:vertex} shows the vertex distribution of candidate events after the $\beta\beta$ selection cuts, and expected $^{214}$Bi background events from a Monte Carlo (MC) simulation for $2.3 < E < 2.7\,{\rm MeV}$. Considering the z-asymmetry of the $^{214}$Bi distribution, the volume is divided into radial-equal-volume bins, 20 bins in the upper and lower hemisphere each for signal-to-background optimization. Due to the larger $^{214}$Bi background on the mini-balloon, the volume bins away from the balloon are expected to have a higher sensitivity, therefore, the background estimation around the central region is especially-important. For the $^{214}$Bi background, the vertex dispersion model was constructed from a full MC simulation based on \texttt{Geant4}~\cite{Agostinelli2003,Allison2006} including decay particle tracking, scintillation photon process, and finite PMT timing resolution. This MC reproduces the observed vertex distance between $^{214}$Bi and $^{214}$Po sequential decay events from the initial radon contamination.

The muon spallation backgrounds come mainly from $^{10}$C, as well as other shorter-lived products, e.g., $^{6}$He, $^{12}$B, and $^{8}$Li. In the phase-2 data, additional event selection criteria to reject the spallation backgrounds are newly introduced based on muon-induced neutron events. Post-muon neutrons are identified by neutron-capture $\gamma$-rays by newly introduced dead-time free electronics (MoGURA), and spherical volume cuts ($\Delta R < 1.6\,{\rm m}$) around the reconstructed neutron vertices are applied for 180\,s after the muon producing the neutrons. In the energy range of the $^{10}$C background ($2.2 < E < 3.5\,{\rm MeV}$), 6 events are rejected within a radius of 1.0 m,  this rate is consistent with the expectation for the LS from a previous study~\cite{Abe2010}. The livetime reduction by this spallation cut is only 7\%.

\section{Results}

Preliminary results presented here are based on the phase-2 data, collected between December 11, 2013, and May 1, 2014, after the $^{110m}$Ag background reduction. The total livetime is 114.8 days.  The livetime, fiducial Xe-LS mass, Xe concentration, $^{136}$Xe mass, and exposure for the data sets in phase-1~\cite{Gando2013} and phase-2 are summarized in Table~\ref{table:fiducial}.

\begin{table}[t]
\begin{tabular}{@{}*{6}{lccccc}}
\hline
\hspace{3.5cm} & \multicolumn{3}{c}{Phase-1~\cite{Gando2013}} & \multicolumn{2}{c}{\hspace{0.5cm} Phase-2} \\ 
\hspace{3.5cm} & ~~~~~\mbox{DS-1}~~~~~ & ~~~~~\mbox{DS-2}~~~~~ & ~~~~Total~~~~ & \hspace{0.5cm} $R < 1.0\,{\rm m}$ & Full Xe-LS \\
\hline
livetime (days) & 112.3 & 101.1 & 213.4 & \hspace{0.5cm}  114.8 & 114.8 \\
fiducial Xe-LS mass (ton) & 8.04 & 5.55 & - & \hspace{0.5cm}  3.27 & 12.88 \\
Xe concentration (wt\%) & 2.44 & 2.48 & - & \hspace{0.5cm}  2.96 & 2.96 \\
$^{136}$Xe mass (kg) & 179 & 125 & - & \hspace{0.5cm}  87.8 & 346 \\
$^{136}$Xe exposure (kg-yr) & 54.9 & 34.6 & 89.5 & \hspace{0.5cm}  27.6 & 108.8 \\
\hline
\end{tabular}
\caption{\label{table:fiducial}Summary of the phase-1 and phase-2 data used in $^{136}$Xe $\beta\beta$ decay analyses.}
\vspace{-0.5cm}
\end{table}

\subsection{Preliminary $2\nu\beta\beta$ analysis}

\begin{figure}
  \includegraphics[width=0.65\columnwidth]{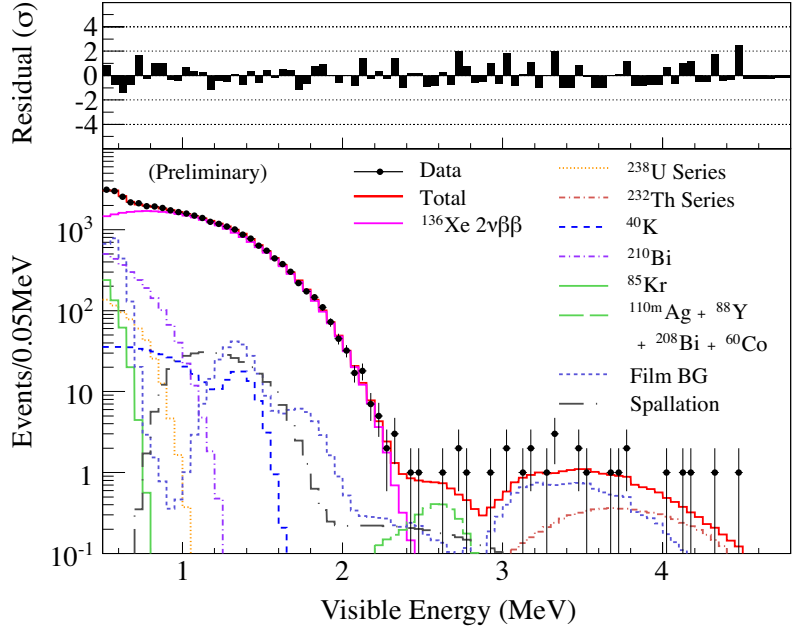}
  \caption{Preliminary energy spectrum of selected $\beta\beta$ candidates within 1.0\,m fiducial radius is shown together with the best-fit backgrounds, with the $2\nu\beta\beta$ decay fit. The residuals from the best-fit are shown in the upper panel.}
  \label{figure:energy_2nu}
\end{figure}

The analysis for the $2\nu\beta\beta$ decay is limited to the volume within the radius of 1.0 m in order to avoid a large $^{134}$Cs/$^{137}$Cs background at the mini-balloon. Fig.~\ref{figure:energy_2nu} shows the energy spectrum of $\beta\beta$ candidates, with a spectral fit, including backgrounds. The measured $2\nu\beta\beta$ decay half-life of $^{136}$Xe is $T_{1/2}^{2\nu} = 2.32 \pm 0.05({\rm stat}) \pm 0.08({\rm syst}) \times 10^{21}$~yr. This is consistent with the previous result based on the phase-1 data, $T_{1/2}^{2\nu} = 2.30 \pm 0.02({\rm stat}) \pm 0.12({\rm syst}) \times 10^{21}$~yr~\cite{Gando2012b}, and with the result obtained by \mbox{EXO-200}, $T_{1/2}^{2\nu} = 2.165 \pm 0.016({\rm stat}) \pm 0.059({\rm syst}) \times 10^{21}$~yr~\cite{Albert2014a}.

\subsection{Preliminary $0\nu\beta\beta$ analysis}

\begin{figure}
  \includegraphics[width=0.45\columnwidth]{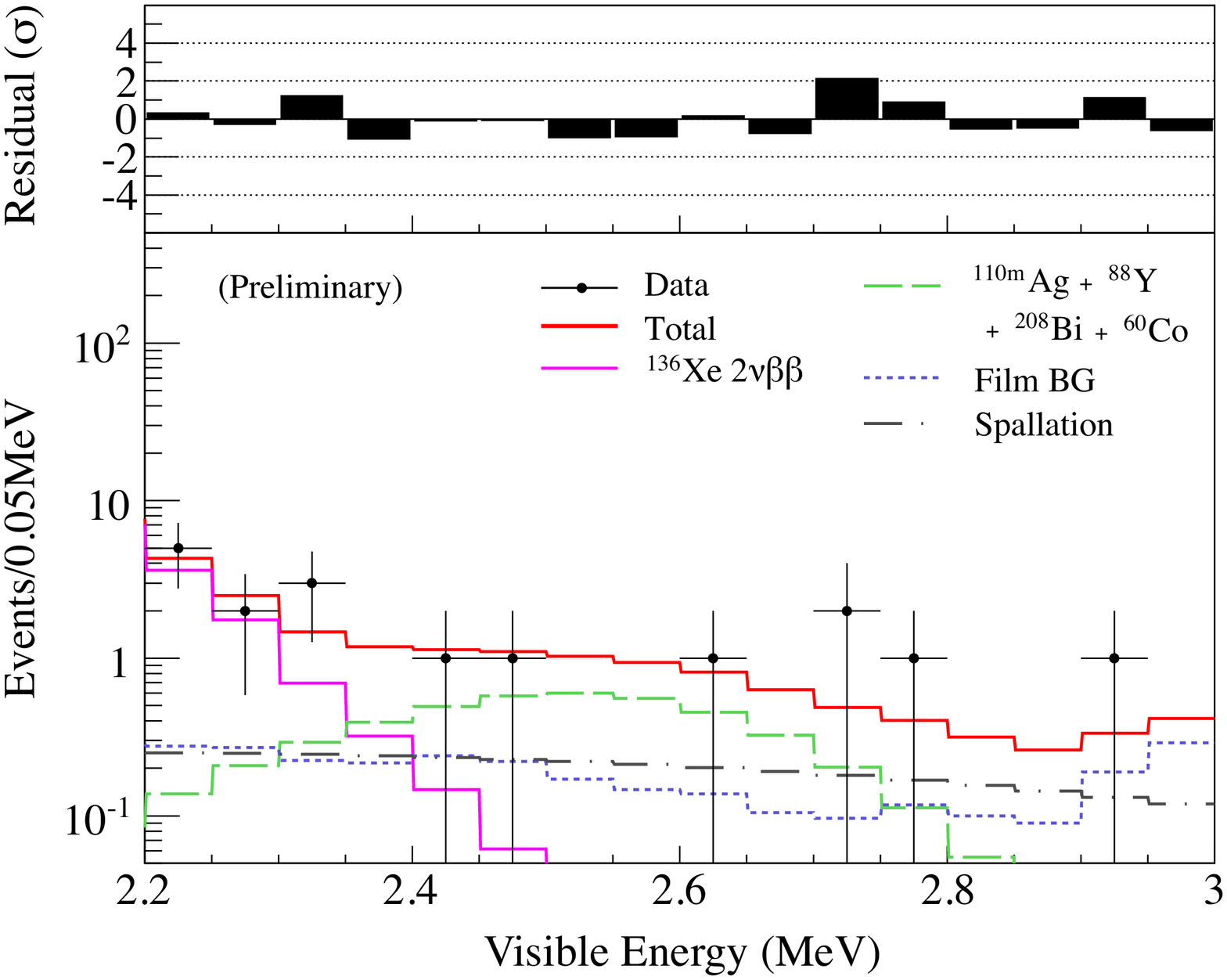}
  \hspace{1.0cm}
  \includegraphics[width=0.45\columnwidth]{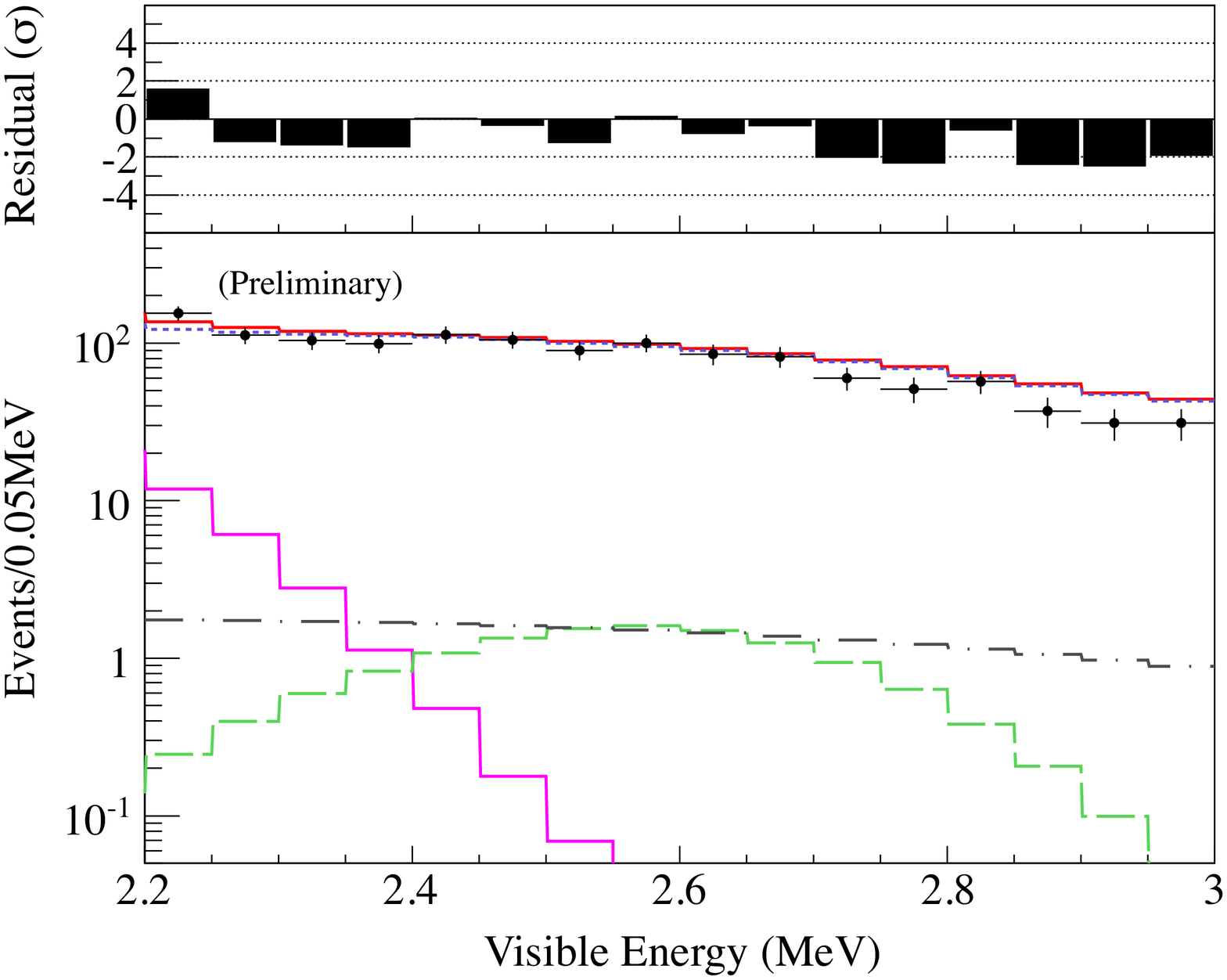}
  \caption{Preliminary energy spectra of selected $\beta\beta$ candidates within the radius cuts, $R < 1.0\,{\rm m}$ (left) and $1.0 < R < 2.0\,{\rm m}$ (right). The best-fit spectra correspond to the 2-dimensional energy-volume analysis fit results described in the text. The residuals from the best-fit are shown in the upper panels.}
  \label{figure:energy_0nu}
\end{figure}

\begin{figure}
  \includegraphics[width=0.7\columnwidth]{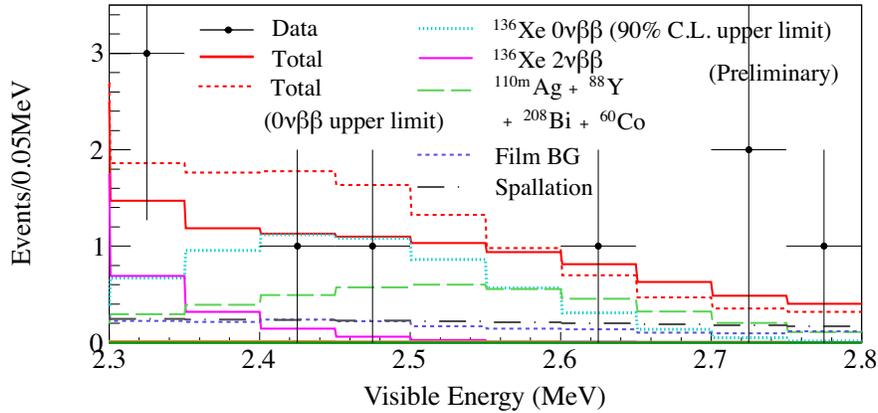}
  \caption{Preliminary energy spectrum of selected $\beta\beta$ candidates within the radius cut $R < 1.0\,{\rm m}$ is shown together with the best-fit backgrounds and the 90\% C.L. upper limit for $0\nu\beta\beta$ decays. This figure shows the data and backgrounds in a narrower range and linear-scale compared to Fig.~\ref{figure:energy_0nu} (left).}
  \label{figure:energy_0nu_limit}
\end{figure}

\begin{figure}
  \includegraphics[width=0.45\columnwidth]{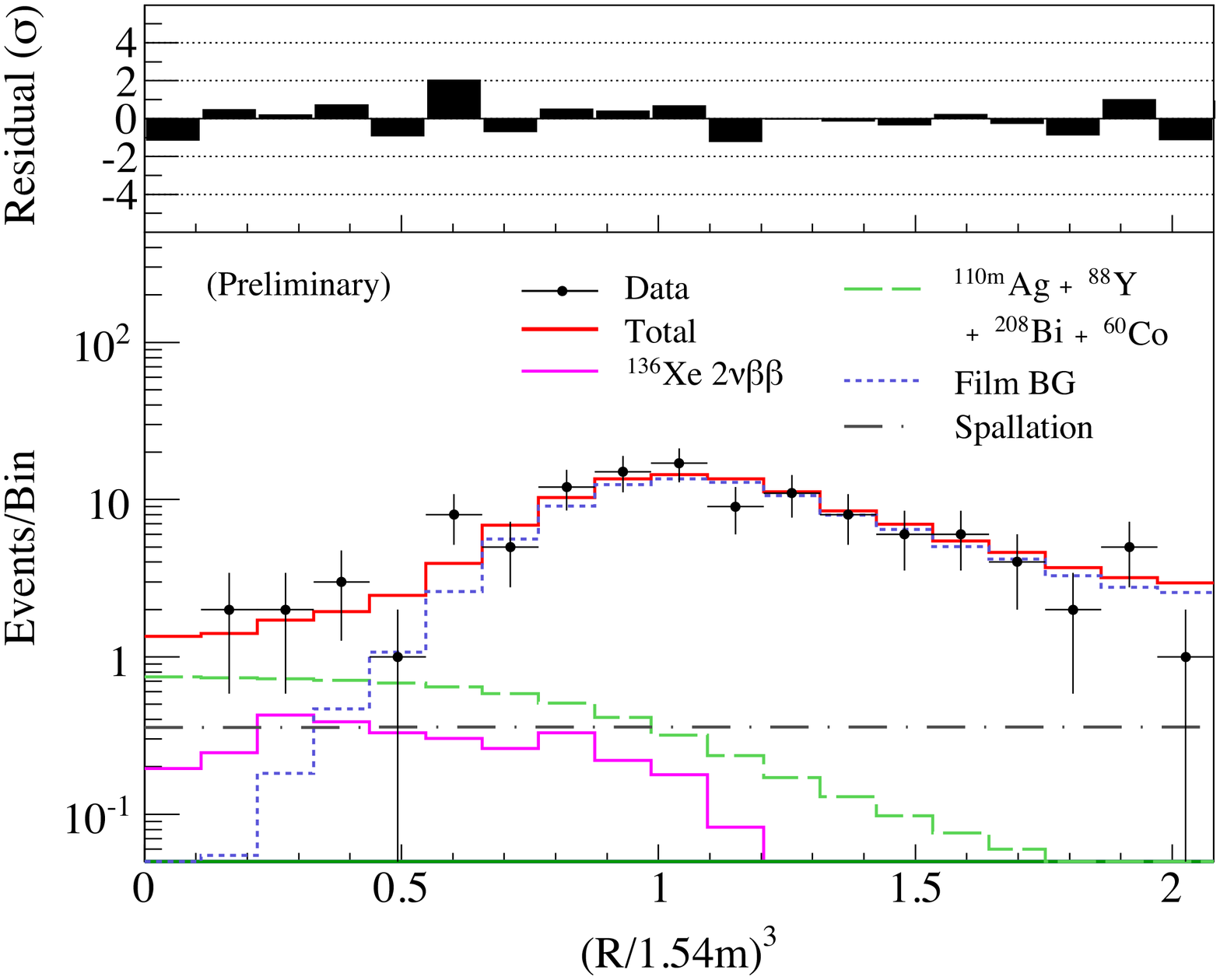}
    \hspace{1.0cm}
  \includegraphics[width=0.45\columnwidth]{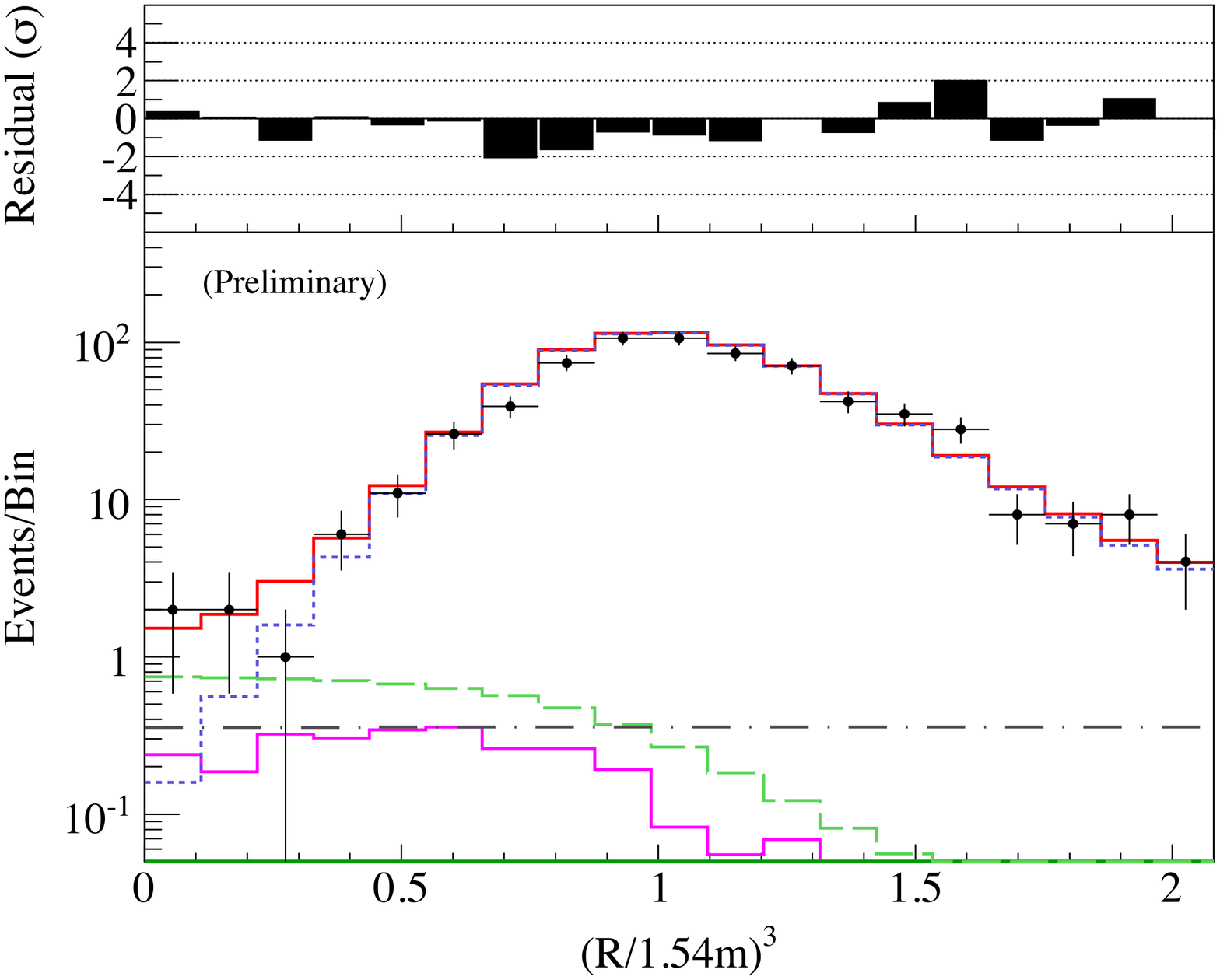}
  \caption{Radius-cube ($R^{3}$) distributions of selected $\beta\beta$ candidates with $2.3 < E < 2.7\,{\rm MeV}$ in the upper hemisphere (left) and the lower hemisphere (right). The radial position was normalized to the mini-balloon radius ($R = 1.54\,{\rm m}$). The backgrounds from the mini-balloon (Film BG) are radially attenuated.}
  \label{figure:R3_0nu}
\end{figure}

The $0\nu\beta\beta$ decay rate is estimated by the fit to 2-dimensional spectra in energy-volume in the full 2-m-radius analysis volume as described above. To simplify the display of the fit results, the energy spectra in only two volumes, the internal volume ($R < 1.0\,{\rm m}$) and the external volume ($1.0 < R < 2.0\,{\rm m}$), are shown in Fig.~\ref{figure:energy_0nu}, together with the best-fit background composition. The potential background contributions of $^{110m}$Ag, $^{88}$Y, $^{208}$Bi, and $^{60}$Co in the $0\nu\beta\beta$ region of interest, as discussed in \mbox{Ref.~\cite{Gando2012a}}, are allowed to vary in the fit. We found no event excess over the background expectation. The 90\% C.L upper limit on the $^{136}$Xe $0\nu\beta\beta$ decay rate is $<$ 17.0~(kton$\cdot$day)$^{-1}$, in Xe-LS mass units. A MC of an ensemble of experiments assuming the best-fit background spectrum without a $0\nu\beta\beta$ signal indicates a sensitivity of $<$ 16~(kton$\cdot$day)$^{-1}$, and the probability of obtaining a stronger limit is 52\%. The $0\nu\beta\beta$ decay contribution corresponding to the 90\% C.L. upper limit for the internal volume is shown in Fig.~\ref{figure:energy_0nu_limit}. The dominant $^{214}$Bi background at the mini-balloon is radially-attenuated, therefore, the data in the inner volume is more sensitive to $0\nu\beta\beta$ decays, as indicated in Fig.~\ref{figure:R3_0nu}. Considering the Xe concentration in the Xe-LS, we obtain a limit on the $^{136}$Xe $0\nu\beta\beta$ decay half-life of $T_{1/2}^{0\nu} > 1.3 \times 10^{25}$~yr (90\% C.L.).

\begin{figure}
  \includegraphics[width=0.6\columnwidth]{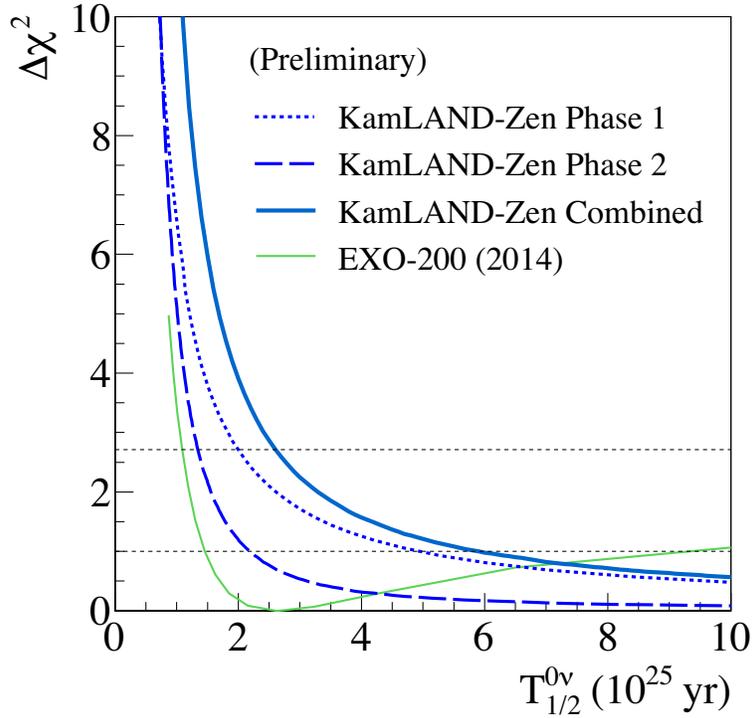}
  \caption{$\Delta\chi^{2}$-profile from the fit to the half-life of $^{136}$Xe $0\nu\beta\beta$ decays in this work (phase-2), the previous work (phase-1), and the combined result (phase-1 $+$ phase-2). The result from \mbox{EXO-200}~\cite{Albert2014b} is also shown for comparison.}
  \label{figure:chi2}
\end{figure}

As shown in Fig.~\ref{figure:chi2}, the combined KamLAND-Zen result from the phase-1~\cite{Gando2013} and phase-2 data gives a 90\% C.L. lower limit of $T_{1/2}^{0\nu} > 2.6 \times 10^{25}$\,yr. This limit is compared to the recent EXO-200 result, which tends to allow shorter half-life values, and gives a 90\% C.L. lower limit of $T_{1/2}^{0\nu} > 1.1 \times 10^{25}$\,yr~\cite{Albert2014b}. Based on nuclear matrix elements (NMEs) from various (R)QRPA models~\cite{Faessler2012}, the combined KamLAND-Zen half-life limit can be converted to a 90\% C.L. upper limit of $\left<m_{\beta\beta}\right> < (140-280)\,{\rm meV}$.

\clearpage

\section{Prospects}

\begin{figure}
  \includegraphics[width=0.65\columnwidth]{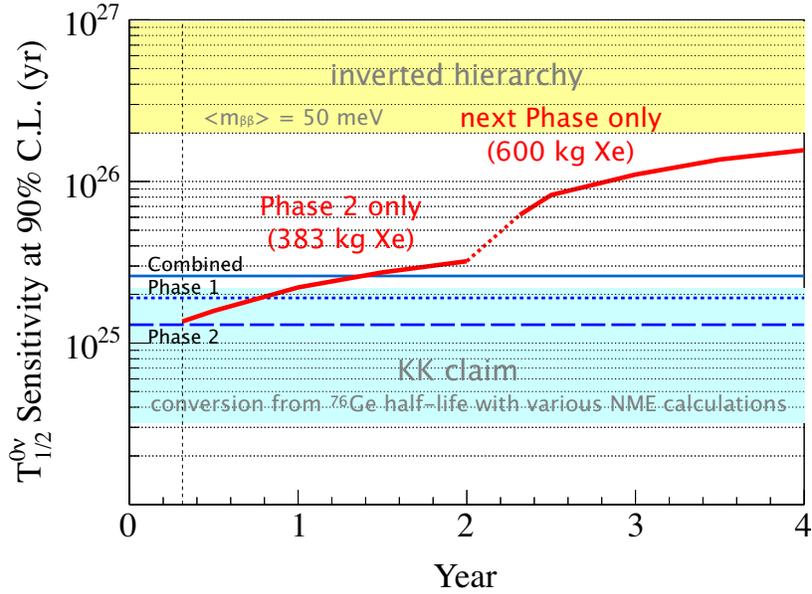}
  \caption{Expected $T_{1/2}^{0\nu}$ sensitivity at 90\% C.L. in the near future for KamLAND-Zen. The red line at less than 2 years corresponds to phase-2 only, and the following red line is next phase only. The three horizontal lines indicate the lower $T_{1/2}^{0\nu}$ limits reported here (phase-2), the previous results (phase-1), and the combined result (phase-1 $+$ phase-2).}
  \label{figure:future}
\end{figure}

The $0\nu\beta\beta$ decay search sensitivity will steadily increase by accumulating additional low background data after the $^{110m}$Ag reduction. Assuming the best-fit background rates in phase-2, the $T_{1/2}^{0\nu}$ sensitivity at 90\% C.L. will reach $3 \times 10^{25}$\,yr within 2 years using the phase-2 data alone, see Fig.~\ref{figure:future}. It will test the claimed observation of  $0\nu\beta\beta$ decay in $^{76}$Ge~\cite{Klapdor2006} more stringently. We plan to rebuild the mini-balloon to increase the Xe amount to 600\,kg (700-800 kg if possible) and reduce the mini-balloon radioactivity by introducing a cleaner material for the balloon film. In that case, owing to the increase of the Xe-LS fiducial mass, the sensitivity will be close to $2 \times 10^{26}$\,yr in a 2 year measurement (Fig.~\ref{figure:future}), which corresponds to $\left<m_{\beta\beta}\right> = 50\,{\rm meV}$ for the largest NME in the (R)QRPA models~\cite{Faessler2012}.

The next near-future $0\nu\beta\beta$ decay search milestone is to reach a sensitivity of $\left<m_{\beta\beta}\right> \sim 20\,{\rm meV}$ which covers the inverted neutrino mass hierarchy. The neutrino mass spectrum may be more clarified by long-baseline neutrino oscillation experiments, cosmological observations, and single-$\beta$ decay experiments in the future. Under such circumstances, the inverted mass hierarchy search will provide an important outcome even without the observation of the positive $0\nu\beta\beta$ signal. A sensitivity covering the inverted hierarchy is projected to be achieved by ``KamLAND2-Zen'', a detector upgrade proposal with better energy resolution against the $2\nu\beta\beta$ background, by introducing light collective mirrors ($1.8 \times$ light yield), new brighter LS ($1.4 \times$ light yield), and high quantum efficiency PMTs ($1.9 \times$ light yield). The energy resolution is expected to be improved from 4.0\% to $<$2.5\% at the Q-value of $^{136}$Xe $\beta\beta$ decay. The enriched Xe amount will be increased to 1,000\,kg or more, and the target sensitivity of 20\,meV will be achieved in a 5 year measurement. The access hole at the top of the detector will be enlarged for the larger mini-balloon installation, as it will also accommodate various additional devices, such as scintillating crystals containing other $0\nu\beta\beta$ decay nuclei, and a NaI crystal for a dark matter search. In addition, other R\&D efforts aiming at the introduction of new technology, such as an imaging device to reject $\gamma$-emitting backgrounds, and scintillating film to reject $^{214}$Bi-$^{214}$Po sequential decay backgrounds, are going forward.

\section{Summary}

KamLAND-Zen realized the initial $0\nu\beta\beta$ decay search by utilizing an extremely low-background detector, and demonstrated the effective background reduction in the xenon loaded liquid scintillator after the purification. We find that the limits on the half-life of $^{136}$Xe $0\nu\beta\beta$ decays and the Majorana neutrino mass are improved. In the near future, the search sensitivity will be enhanced by accumulating additional low background data. A phased-program with several detector improvements is planned for even better sensitivity enhancement.


\begin{theacknowledgments}
The \mbox{KamLAND-Zen} experiment is supported by the Grant-in-Aid for Specially Promoted Research under grant 21000001 of the Japanese Ministry of Education, Culture, Sports, Science and Technology; the World Premier International Research Center Initiative (WPI Initiative), MEXT, Japan; Stichting FOM in the Netherlands; and under the US Department of Energy, Office of Science, Office of Nuclear Physics under contract No. DE-AC02-05CH11231, as well as other DOE awards to individual institutions. The Kamioka Mining and Smelting Company has provided service for activities in the mine.
\end{theacknowledgments}



\bibliographystyle{aipproc}   

\bibliography{KamLAND-Zen-Results}

\IfFileExists{\jobname.bbl}{}
 {\typeout{}
  \typeout{******************************************}
  \typeout{** Please run "bibtex \jobname" to optain}
  \typeout{** the bibliography and then re-run LaTeX}
  \typeout{** twice to fix the references!}
  \typeout{******************************************}
  \typeout{}
 }

\end{document}